\documentclass[varenna]{cimento}

\usepackage{graphicx}
\usepackage{amsfonts,amssymb,amsmath}
\usepackage{color}
\usepackage{eimodels}
\usepackage{eivars}
\usepackage{xspace}
\usepackage{fancyhdr}
\usepackage{bm}

\newcommand{\calP}{\mathcal{P}}

\newcommand{\calO}{\mathcal{O}}

\newcommand{\calK}{\mathcal{K}}
\newcommand{\calL}{\mathcal{L}}

\newcommand{\GeV}{\mathrm{GeV}}

\newcommand{\km}{\mathrm{km}}

\newcommand{\Mpc}{\mathrm{Mpc}}

\newcommand{\Mp}{M_\usssPl}

\usepackage{graphicx}  % got figures? uncomment this
\title{The Theory of Inflation}

\author{J.~Martin\thanks{jmartin@iap.fr}}

\institute{Institut d'Astrophysique de Paris, UMR 7096-CNRS,
  Universit\'e Pierre et Marie Curie, 98 bis boulevard Arago, 75014
  Paris, France}

\begin{document}

\maketitle

\begin{abstract}
  This article contains a concise review of the theory of
  inflation. We discuss its main theoretical aspects as well as its
  observational predictions. We also explain how the most recent
  astrophysical observations constrain the inflationary scenario.
\end{abstract}

\section{Introduction}
\label{sec:intro}

The theory of inflation was invented at the end of the $70$'s and
beginning of the $80$'s in order to improve the hot Big Bang
model~\cite{Starobinsky:1980te,Starobinsky:1982ee,
  Guth:1980zm,Linde:1981mu,Albrecht:1982wi,Linde:1983gd}. It consists
in a phase of accelerated expansion taking place in the early
Universe, at very high energy scales, possibly as high as
$10^{15}\GeV$. Not only inflation solves the puzzles of the standard
model but it also provides a convincing mechanism for structure
formation~\cite{Starobinsky:1979ty,Mukhanov:1981xt,Mukhanov:1982nu,Guth:1982ec,
  Hawking:1982cz,Bardeen:1983qw} (for reviews, see {\textit e.g.}
Refs.~\cite{Martin:2004um,Martin:2007bw}) which, interestingly enough,
is based on General Relativity (GR) and Quantum Mechanics (QM), two
theories notoriously difficult to combine.

On the observational front, the progresses have also been enormous,
culminating recently with the publication of the high accuracy
measurement of the Cosmic Microwave Background (CMB) anisotropies by
the European Space Agency (ESA) satellite
Planck~\cite{Ade:2013zuv,Planck:2013jfk,Ade:2015xua,Ade:2015lrj}. For
the first time, this satellite has been able to show that the spectral
index of the scalar power spectrum is close to one (exact scale
invariance) but not exactly one, the deviation from one being detected
at a statistically significant level, namely at more than
$5\sigma$. This is a crucial landmark because this was a prediction of
inflation (and not a post-diction). This is the reason why inflation is
now viewed as the front-runner candidate for describing the physical
conditions that prevailed in the early Universe~\cite{Martin:2015dha}.

The aim of these lectures is to give a brief introduction to the
theory of inflation. It is organized as follows. In the next section,
Section~\ref{sec:why}, we discuss the motivations for inflation. In
Section~\ref{subsec:premodel}, we first present the standard model of
Cosmology, the hot Big Bang phase, as it was prior to the invention of
inflation. Then, in Section~\ref{subsec:puzzles}, we discuss the
puzzles of the hot Big Bang phase and why a phase of accelerated
expansion can solve them. In Section~\ref{subsec:basics}, we discuss
how inflation can be realized in practice and how it comes to an end
(the theory of reheating). In Section~\ref{sec:perturbations}, we
discuss the theory of inflationary cosmological perturbations of
quantum-mechanical origin. We first show that the quantum state of the
perturbations at the end of inflation is peculiar (a two-mode squeezed
state) and then we calculate the power spectrum in the slow-roll
approximation. In Section~\ref{sec:extensions}, we briefly describe
more complicated ways to realize inflation, in particular multiple
field inflation. In Section~\ref{sec:observations}, we discuss the
observational status of inflation. We argue that the simplest class of
scenarios is the preferred one and present observational constraints
on the shape of the potential and on the reheating phase. Finally, in
Section~\ref{sec:conclusions}, we recap the main points and briefly
discuss the future of inflation.

\section{Why Inflation?}
\label{sec:why}

\subsection{The pre-inflationary standard model}
\label{subsec:premodel}

Among the four fundamental interactions that have been identified in
Nature, gravity is the important one when it comes to
Cosmology. Indeed, the Universe being neutral, this is the only force
left with an infinite range and, therefore, the only one which can
shape the Universe on astrophysical scales. The gravitational
interaction being described by GR, any attempt to construct a model of
the cosmos must be based on this theory. In addition, the standard
model of cosmology, the so-called hot Big Bang model, is based on a
second fundamental assumption, namely the cosmological principle which
states that, on large scales, the Universe in homogeneous and
isotropic. This means that the general relativistic metric describing
our Universe can be taken to be the Friedman-Lemaitre-Robertson-Walker
(FLRW) one, namely
\begin{equation}
{\rm d}s^2=-{\rm d}t^2+a^2(t)\left(\frac{{\rm d}r^2}{1-\calK r^2}
+r^2{\rm d}\theta^2+r^2\sin ^2 \theta {\rm d}\varphi^2\right),
\end{equation}
where $a(t)$ is the scale factor and $\calK$ is a constant related to the
curvature radius of space $r_{\rm curv}=a(t)/\sqrt{\vert
  \calK\vert}$.
Assuming that matter is described by perfect fluids, the corresponding
Einstein equations read
\begin{align}
\label{eq:fried}
\frac{\dot{a}^2}{a^2}+\frac{{\cal K}}{a^2} &=
\frac{1}{3\Mp^2}\sum _{i=1}^{N}\rho _i+\frac{\Lambda _{_{\rm B}}}{3}\, , 
\\
-\biggl(2\frac{\ddot{a}}{a}+\frac{\dot{a}^2}{a^2}
+\frac{{\cal K}}{a^2}\biggr) 
& =\frac{1}{\Mp^2} \sum _{i=1}^Np_i-\Lambda_{_{\rm B}} \, ,
\end{align}
where $\Mp$ is the Planck mass and $\Lambda_{_{\rm B}}$ is the bared
cosmological constant. The quantities $\rho_i$ and $p_i$ are
respectively the energy density and pressure of the fluid ``$i$''. In
the hot Big Bang model, one has five species, photons, neutrinos
(which form radiation) and cold dark matter (cdm) and baryons (which
form pressure-less matter) plus dark energy (given by the cosmological
constant). Photons and neutrinos have a constant equation of state
equals to $1/3$, which means that $p_\gamma=\rho_\gamma/3$ and
$p_\nu=\rho_\nu/3$. As already mentioned cdm and baryons have
vanishing pressure. Finally dark energy (de) has a vacuum equation of
state, meaning that $p_{\rm de}=-\rho_{\rm de}$. The standard model is
also such that the spatial curvature vanishes, $\calK =0$. The free
parameters are $H_0\equiv \dot{a}/a\vert_{\rm now}$ (a dot denotes a
derivative with respect to cosmic time), $\Lambda _{_{\rm B}}$,
$\rho_\gamma$, $\rho_\nu$, $\rho_{\rm cdm}$, $\rho_{\rm b}$ and $\tau$
the optical depth that describes how the universe re-ionizes. We also
have two extra parameters describing the perturbations, $A_{_{\rm S}}$
and $\nS$ that will be introduced later on. This means a total of nine
parameters. However, introducing the critical energy density
$\rho_{\rm cri}=3H^2\Mp^2$ and defining
$\Omega_i\equiv \rho_i/\rho_{\rm cri}$, the fact that $\calK =0$ means
that the Friedman equation~(\ref{eq:fried}) can be rewritten as a
constraint,
$\Omega_\gamma+\Omega_\nu+\Omega_{\rm cdm}+\Omega_{\rm b}+\Omega_{\rm
  de}\equiv \Omega_{\rm tot}=1$.
So, in fact, we have eight free parameters (often, $\rho_\gamma$ and
$\rho_\nu$ are not viewed as free parameters because they are
precisely determined by the CMB measurement and the number of
neutrinos family; in that case we have a six parameter model). These
free parameters have now been measured with good precision (at the
percentage level)~\cite{Ade:2013zuv,Ade:2015xua}. For the expansion
rate, one has $H_0=100 h \, \km \times \mbox{s}^{-1}\times \Mpc^{-1}$
with $h\simeq 0.67$, and for the matter content in the present day
Universe, $\Omega_\gamma h^2\simeq 2.47\times 10^{-5}$,
$\Omega _\nu h^2\simeq 1.68\times 10^{-5}$ (assuming three families of
neutrinos), $\Omega_{\rm cdm} h^2\simeq 0.1198$,
$\Omega_{\rm b}h^2\simeq 0.02255$ and
$\Omega_{\rm de}h^2\simeq 0.306$.

Knowing the matter content, by integrating the Einstein equations, we
can infer the history of the Universe. The early Universe was
dominated by radiation, with a scale factor given by
$a(t)\propto t^{1/2}$ from the initial singularity until a redshift
$z_{\rm eq}\simeq 3400$. Then, pressure-less matter took over with a
scale factor $a(t)\propto t^{2/3}$ until a redshift of order
one. Then, dark energy started to dominate and we still live in this
epoch. The history of the Universe is thus made of three successive
eras. 

This simple model, except for the presence of dark energy, was already
known before the $80$'s (although, at that time, the parameters were
not measured with today accuracy) and has a great explanatory
power. As mentioned before, it is known as the hot Big Bang model or
the $\Lambda \mbox{CDM}$ model in its most modern incarnation and is
considered as the most convincing model for cosmology. Why, then, the
simple version presented above is nevertheless considered as not fully
satisfactory thus motivating the introduction of inflation? We now turn
to this question.

\subsection{The puzzles of the standard model}
\label{subsec:puzzles}

With a few parameters, the pre-inflationary standard model of
Cosmology was (is) able to explain a very large number of
observational facts. Therefore it may seem strange to view it as not
totally satisfactory. In fact, the difficulties of the hot Big Bang
model are all related to the initial conditions. For instance, it is
difficult to understand why spatial curvature is so small
today. Indeed, the expansion during the hot Big Bang phase is
decelerated and this means that $\Omega_{\rm tot}-1$ is
growing. Therefore, since $\Omega_{\rm tot}-1$ is, today, very close
to zero, this implies that it was in fact incredibly small in the
early Universe (say, at BBN). Of course, it is always possible to
postulate that the initial conditions were just such that it was the
case. However, there is another explanation which consists in assuming
that there was an accelerated phase of expansion, $\ddot{a}>0$, prior
to the hot Big Bang epoch. This new phase of accelerated expansion is
called ``inflation''. Then, the initial conditions at the beginning of
the hot Big Bang epoch are now viewed as the ``final conditions'' at
the end of inflation. Moreover, during a phase of accelerated
expansion $\Omega_{\rm tot}-1$ is decreasing. Therefore, if
$\Omega_{\rm tot}-1$ sufficiently decreases during inflation, it can
entirely compensate the subsequent growth during the hot Big Bang
phase and we understand why it is still small today. One can show that
the compensation occurs if we have more than $60$ e-folds of
inflation. In some sense, inflation is a physical mechanism which puts
the hot Big Bang phase on the ``right tracks'' by automatically
single outing the right initial conditions.

Quite remarkably, one can show that all the puzzles of the standard
model can be solved by the same mechanism~\cite{Guth:1980zm}. For
instance, this is the case of the so-called horizon problem. According
to the hot Big Bang model, the angular scale of the horizon on the
last scattering surface (where the CMB radiation was emitted) is
$\simeq 1^{\circ}$. This means that we should expect the temperature
to be strongly inhomogeneous on this scale all over the sky. As is
well-known, this is not the case since the CMB is, on the contrary,
extremely homogeneous and isotropic. However, if one has $60$ e-folds
of inflation before the hot Big Bang phase, then the horizon at the
last scattering surface covers the entire celestial sphere today and
the problem is gone. We stress again that the number of e-folds needed
to solve the problem turns out to be the same as for the flatness
problem, namely $60$.

Of course, postulating a phase of accelerating is not sufficient. One
must also identify a physical mechanism that could be responsible for
it. In the next section, we discuss this question.

\subsection{Basics of inflation}
\label{subsec:basics}

We have seen before that, if there is a phase of accelerated expansion
in the early Universe, then the puzzles of the hot Big Bang model can
be explained. As long as the gravitational field is described by GR
and the cosmological principle valid, the acceleration of the scale
factor can be expressed as
\begin{equation}
\label{eq:accela}
\frac{\ddot{a}}{a}=-\frac{1}{6\Mp^2}\sum _{i=1}^N(\rho _i
+3p_i)+\frac{1}{3}\Lambda_{_{\rm B}}  .
\end{equation}
Assuming that the cosmological constant does not play a role in the
early Universe (given its present day value), the condition for having
$\ddot{a}>0$ reads
\begin{equation}
\label{eq:condddapositive}
\rho _{_{\rm T}}+3p_{_{\rm T}}<0 \, ,
\end{equation}
where $\rho_{_{\rm T}}=\sum_{i=1}^N\rho_i $ and
$p_{_{\rm T}}=\sum_{i=1}^Np_i $ denote the total energy density and
pressure. Given that the energy density must be positive, we are left
with the condition that the pressure must be negative.

In usual situations, the pressure of a fluid is positive. This is for
instance the case of radiation. However, inflation is supposed to take
place in the very early Universe, at extremely high redshifts, and at
those energies, hydrodynamics is clearly not the appropriate framework
to describe matter. We should rather use field theory. The simplest
type of field, compatible with the cosmological principle and the FLRW
symmetries is a scalar field. We therefore assume that the matter
content of the early Universe was dominated by a homogeneous scalar
field $\phi(t)$ called, for obvious reasons, the ``inflaton''. The
corresponding action is given by
\begin{equation}
\label{eq:normallagrange}
  \calL=-\frac{1}{2}g^{\mu \nu}\partial_{\mu}\phi\partial_{\nu }\phi
-V(\phi)+\calL_{\rm int}(\phi,A_{\mu},\Psi),
\end{equation}
where $V(\phi)$ is the inflaton potential and $\calL_{\rm int}$
describes the interaction of the inflaton field with the other fields
present such as gauge bosons $A_{\mu}$ or fermions $\Psi$. Then, by
varying this action with respect to the metric tensor, one can
calculate the energy momentum tensor and, therefore, the energy density
and the pressure of the system. Ignoring for the moment the
interaction term, this leads to
\begin{equation}
\rho =\frac{\dot{\phi}^2}{2}+V(\phi), \quad 
 p=\frac{\dot{\phi}^2}{2}-V(\phi).
\end{equation}
We see that energy density is positive definite as it should [of
course, $V(\phi)>0$] but this is not the case of pressure. If the
potential energy dominates over the kinetic energy, then $p<0$. This
will be the case if the kinetic energy is small or, in other words, if
the inflaton moves slowly along its potential. And this will happen if
the potential is nearly flat. We conclude that, if the inflaton
dominates the energy budget at early times and if its potential is
almost flat, then a phase of inflation can occur. This is the basics
idea that underlies the theory of inflation.

At the technical level, the evolution of the system is
controlled by the Friedmann and Klein-Gordon equations, namely
\begin{equation}
\label{eq:friedman/kg}
H^2 =\frac{1}{3\Mp^2}\left[\frac{\dot{\phi}^2}{2}+V(\phi)\right], \quad
\ddot{\phi}  +3H\dot{\phi}+V_{\phi} = 0,
\end{equation}
where a subscript $\phi$ means a derivative with respect to the
inflaton field. Unfortunately, this system of equations cannot be
solved analytically unless the potential has a very specific form [for
instance, $V(\phi)\propto e^{-\alpha \phi}$, a model called power law
inflation]. Therefore, we have to use either numerical calculations or
a perturbative method. In general, a perturbative method is based on
an expansion of the relevant physical quantities in terms of a small
parameter (or several) naturally present in the problem (for instance
a coupling constant in field theory). Here, one can use the fact that
the potential is nearly flat. If it is exactly flat, then the scalar
field acts as a cosmological constant and the corresponding solution
is de Sitter. One can then expand the solution of the
system~(\ref{eq:friedman/kg}) around de Sitter. Since the de Sitter
solution corresponds to a constant Hubble parameter, one can define
small parameters by considering the derivatives of $H$ and, then,
expand the solution in these parameters. They are called horizon flow
parameters or slow-roll parameters and are defined
by~\cite{Schwarz:2001vv,Leach:2002ar}
\begin{equation}
\label{eq:defhf}
\epsilon_{n+1} \equiv \frac{{\rm d} \ln \left \vert 
\epsilon_n \right \vert}{{\rm d} N}, 
\quad n\ge 0,
\end{equation}
where $\epsilon_0\equiv H_\uini/H$ stands at the top of the hierarchy
and $N\equiv \ln(a/a_\uini)$ is the number of e-folds. The first
Hubble flow parameter can be expressed as
\begin{equation}
\label{eq:eps1}
\epsilon_1=-\frac{\dot{H}}{H^2}=1-\frac{\ddot{a}}{aH^2}
=\frac{3\dot{\phi}^2}{2}\frac{1}{\dot{\phi}^2/2+V(\phi)}.
\end{equation}
As mentioned above, it is related to the first derivative of the
Hubble parameter. The second Hubble flow parameter, $\epsilon_2$,
would be related to $\ddot{H}$ and so on. We also see on the second
expression of $\epsilon_1$ that $\epsilon_1<1$ when $\ddot{a}>0$, that
is to say when inflation occurs. Of course, $\epsilon_1\ll 1$ when the
inflationary expansion is close to that of de Sitter. Finally, the
third expression of $\epsilon_1$ makes clear that it is a very small
quantity when the kinetic energy is small compared to the total energy
and, therefore, compared to the potential energy. In fact, there is
yet another way to express the Hubble flow parameters. If one assumes
that $\epsilon_n\ll 1$ [the following expressions are therefore
approximate contrary to Eqs.~(\ref{eq:eps1}) which are exact], then
the first three Hubble flow parameters can be written as
as~\cite{Liddle:1994dx}
\begin{align} 
\label{eq:epsfirst}
\epsilon_1 & \simeq
\frac{\Mp^2}{2}\left(\frac{V_\phi}{V}\right)^2, \\ 
\label{eq:eps2}
\epsilon_2 & \simeq
2\Mp^2\left[\left(\frac{V_\phi}{V}\right)^2
-\frac{V_{\phi \phi}}{V}\right], \\
\label{eq:eps3}
\epsilon_2\epsilon_3 & \simeq 2\Mp^4\left[
\frac{V_{\phi \phi \phi}V_\phi}{V^2}-3\frac{V_{\phi \phi}}{V}
\left(\frac{V_\phi}{V}\right)^2
+2\left(\frac{V_\phi}{V}\right)^4\right].
\end{align} 
It is then clear that, when the inflaton potential is nearly flat, one
has $\epsilon_n\ll 1$. The Hubble flow parameters are in fact nothing
but a measure of the flatness of the inflaton potential.

Having identified the small parameters of the problem, one can now use
them and design a method of approximation based on an expansion in
terms of the $\epsilon_n$'s. This is called the slow-roll
approximation. The first step consists in re-writing the Friedman and
Klein-Gordon equations~(\ref{eq:friedman/kg}) in terms of the
$\epsilon_n$'s. This leads to
\begin{eqnarray}
\label{eq:exactfried}
H^2 &=& \frac{V}{\Mp^2(3-\epsilon_1)}, \\
\left(1 + \dfrac{\epsilon_2}{6 - 2\epsilon_1}\right)
\dfrac{\ud \phi}{\ud N} &=& - \Mp^2 \dfrac{\ud \ln V}{\ud \phi}\,.
\end{eqnarray}
At this stage, these expressions are exact. Then, we expand them at
leading order in the Hubble flow parameters. This gives
\begin{equation}
H^2\simeq \frac{V}{3\Mp^2}, \quad 
\frac{{\rm d}\phi}{{\rm d}N}\simeq -\Mp^2\frac{{\rm d}\ln V}{{\rm d}\phi}.
\end{equation}
Unsurprisingly, we now see that the expansion rate of the Universe is
solely controlled by the potential energy. One great advantage of the
above equations is that they can be integrated exactly. The solution
reads
\begin{equation}
\label{eq:srtrajectory}
N-\Nini=-\frac{1}{\Mp^2}\int_{\phiini}^{\phi}\frac{V(\chi)}
{V_\chi(\chi)}\, \ud \chi \, ,
\end{equation} 
$\phiini$ being the initial value of the inflaton. If the above
integral can be performed, then one obtains $N=N(\phi)$ and by
inverting it, one arrives at the trajectory, $\phi=\phi(N)$. If one
assumes a potential $V(\phi)$, this solution can be compared with the
exact solution obtained by a numerical integration. In practice, as
long as $\epsilon_n\ll 1$, Eq.~(\ref{eq:srtrajectory}) turns out to be
an excellent approximation.

We now turn to another crucial question of the inflationary scenario,
namely how it comes to an
end~\cite{Turner:1983he,Traschen:1990sw,Kofman:1997yn,Amin:2014eta}. At
this stage, let us recall that inflation is not an alternative to the
$\Lambda$CDM model but just an additional ingredient. A phase of
inflation is supposed to take place in the early Universe for the
reasons explained in Sec.~\ref{subsec:puzzles} but, then, it must be
smoothly connected to the standard $\Lambda$CDM phase. On a more
practical side, it is known that the expansion of the Universe was
radiation dominated during the Big Bang Nucleosynthesis (BBN)
(otherwise the production of light elements, which is known to be in
good agreement with the data, would be drastically modified) and,
therefore, inflation must have stopped by that time.

There exists different mechanisms to stop inflation but the simplest
one is just that, at some point, the potential is no longer flat
enough to support inflation. Usually this happens in the vicinity of
the minimum of the potential. Technically, this means that the
slow-roll approximation is no longer valid. In fact, from
Eq.~(\ref{eq:eps1}), one sees that the expansion is no longer
accelerated when $\epsilon_1=1$ which, therefore, defines the time at
which inflation comes to an end. Then, the field starts oscillating at
the bottom of its potential. If $m^2={\rm d}^2V/{\rm d}\phi^2$ is the
mass around the local minimum, the field behaves
as~\cite{Turner:1983he}
\begin{equation}
\label{eq:inflatonosci}
\phi(t)=\phi_{\rm end}\left(\frac{a_{\rm end}}{a}\right)^{3/2}
\sin \left(mt\right),
\end{equation}
namely the field oscillates with a frequency given by its mass. Of
course, in this regime, the kinetic energy is no longer sub-dominant
compared to the potential energy. In fact, there is now equipartition
between them which means that $\left \langle p\right \rangle_t=0$.
This implies that the averaged energy density behaves as dust as also
revealed by the fact that the overall amplitude of the inflaton is
proportional to $a^{-3/2}$.

The above behavior is valid if one neglects the interaction of the
inflaton with the other fields or, in other words, for times much
smaller than the inflaton life time $\Gamma^{-1}$, where $\Gamma $ is
the total inflaton decay rate. If this is taken into account, then
Eq.~(\ref{eq:inflatonosci}) becomes
\begin{equation}
\label{eq:inflatonoscigamma}
\phi(t)=\phi_{\rm end}e^{-\Gamma t}\left(\frac{a_{\rm end}}{a}\right)^{3/2}
\sin \left(mt\right),
\end{equation}
which shows that the total energy density stored in the inflaton field
quickly goes to zero. This energy is transferred to the inflaton decay
products. Then, these decay products thermalize and the radiation
dominated epoch starts at a temperature which is known as the
reheating temperature $T_{\rm rh}$. This is the first time that a
temperature can be defined in the history of the
Universe. Equivalently, this also determines the reheating energy
density, $\rho_{\rm reh}$, that is to say the energy density at which
one starts the $\Lambda$CDM model. It is given by
\begin{align}
\rho_{\rm reh}=g_*\frac{\pi^2}{30}T_{\rm reh}^4,
\end{align}
where $g_*$ encodes the number of relativistic degrees of freedom. 

It is also interesting to study the evolution of the equation of state
during the reheating. We know it must transit between $-1$ and
$1/3$. In fact, observationally speaking, the mean equation of state
is easier to probe. It is defined
by~\cite{Martin:2006rs,Martin:2010kz,Martin:2014nya,Martin:2016oyk}
\begin{equation} 
\label{eq:wrehbar}
\wrehbar \equiv \frac{1}{\Delta N}\int_{\Nend}^{\Nreh} \wreh(n){\rm d} n,
\end{equation}
where $\Delta N \equiv \Nreh - \Nend$ is the total number of e-folds
during reheating and $\wreh\equiv p_{_{\rm T}}/\rho_{_{\rm T}}$ is the
instantaneous equation of state. The quantity $\wrehbar$ controls the
evolution of the total energy density since one has
\begin{equation}
  \rho_{\rm reh}=\rho_{\rm end}\, e^{-3(1+\wrehbar)\Delta N},
\end{equation} 
where $\rho_{\rm end}$ is the energy density at the end of inflation,
namely when $\epsilon_1=1$. If one is given a model of inflation, then
this quantity can be easily calculated.

It is also relevant to introduce the reheating parameter which is a
quantity depending on $\rho_{\rm reh}$ and $\wrehbar$. Explicitly, it
reads~\cite{Martin:2006rs,Martin:2010kz,Martin:2014nya,Martin:2016oyk}
\begin{equation}
\label{eq:defRrad}
\Rrad\equiv 
\left(\frac{\rho_{\rm reh}}
{\rho_{\rm end}}\right)^{(1-3\wrehbar)/(12+12\wrehbar)}.
\end{equation}
The reason why this parameter is important can be found in
Refs.~\cite{Martin:2006rs,Martin:2010kz,Martin:2014nya,Martin:2015dha,Martin:2016oyk}. It
turns out that, when one tries to constrain reheating with the CMB,
we end up constraining this parameter. As simple check allows us to
understand why. Observationally speaking there should not be any
difference between a model where reheating proceeds instantaneously
and a model where reheating proceeds with an equation of state
$1/3$. If $\Rrad$ is the only combination of parameters we can access
to, it should therefore have the same value for those two
situations. And, indeed, it is easy to check that $\Rrad=1$ if
$\rho_{\rm reh}=\rho_{\rm end}$ (instantaneous reheating) or
$\wrehbar=1/3$ (radiative reheating).

Let us now illustrate the previous considerations on a simple
example. Supposed the inflaton potential is given by
$V(\phi)=m^2\phi^2/2$. Then, it is easy to perform the integral 
in Eq.~(\ref{eq:srtrajectory}) and the corresponding trajectory 
reads
\begin{equation}
\phi(N)=\sqrt{\phi_{\rm ini}^2-4\Mp^2(N-N_{\rm ini})}
\end{equation}
As explained before, inflation stops when $\epsilon_1=1$ which, in
this case, means $\phi_{\rm end}=\sqrt{2}\Mp$. From this result, one
can also compute the total number of e-folds. One finds
\begin{equation}
N_{_{\rm T}}\equiv N_{\rm end}-N_{\rm ini}=\frac14 \frac{\phi_{\rm ini}^2}{\Mp^2}
-\frac12.
\end{equation}
This relation means that, in order to have more than $60$ e-folds, one
should start from $\phi_{\rm ini}\gtrsim 15\Mp$. Finally, the
reheating will be completed when $H\simeq \Gamma$, namely
$g_*\pi^2T_{\rm reh}^4/30\simeq \Mp^2\Gamma^2$ or
\begin{equation}
T_{\rm reh}\simeq \left(\frac{30}{g_*\pi^2}\right)^{1/4}\Mp^{1/2}\Gamma ^{1/2}.
\end{equation}
We see that the reheating temperature scales as the square root of the
decay rate.

\section{Inflationary Cosmological Perturbations}
\label{sec:perturbations}

We now turn to the theory of cosmological perturbations of quantum
mechanical origin. This part of the inflationary scenario makes use of
GR and QM and as such is particularly interesting. Moreover, it allows
us to build a bridge between theoretical considerations and actual
astrophysical measurements. Therefore, it plays a crucial role in our
attempts to observationally probe inflation.

So far, we have considered that the Universe was homogeneous and
isotropic. Clearly, in the real world, this is not the case. Going
beyond the cosmological principle is a priori technically challenging
since this means solving Einstein equations in an inhomogeneous and
anisotropic situation. Fortunately, we know that the amplitude of
these inhomogeneities were small in the early Universe as revealed by
the fact that $\delta T/T\simeq 10^{-5}$ on the last scattering
surface located at a redshift of $z_{\rm lss}\simeq 1100$. Since the
amplification of the fluctuations proceeds by gravitational collapse,
the amplitude of the inhomogeneities were even smaller during
inflation. As a consequence, one can study their behavior
perturbatively. Moreover, restricting ourselves to linear
perturbations (leading order) is sufficient. Based on the previous
considerations, we can then write~\cite{Mukhanov:1990me}
\begin{equation}
  g_{\mu \nu}(\eta,{\bm x})=g_{\mu \nu}^{\rm FLRW}(t)+\delta g_{\mu \nu}
\left(\eta,{\bm x}\right)+\cdots
\end{equation}
with the assumption that
$\left \vert \delta g_{\mu \nu}(\eta,{\bm x}) \right \vert \ll \left
  \vert g_{\mu \nu}^{\rm FLRW}(t)\right \vert$.
The tensor $\delta g_{\mu \nu}$ can be decomposed in three types of
fluctuations, scalar, vector and tensor or gravitational waves. The
study of scalar perturbations can be reduced to the study of a single
quantity, the curvature perturbation $\zeta(\eta,{\bm x})$ and the
primordial gravitational waves can be described by a transverse and
traceless two rank tensor $h_{ij}(\eta,{\bm x})$,
$h^i{}_i=\partial _ih^i{}_j=0$. Vector perturbations do not play a
role during inflation. As was already mentioned, the evolution of the
Universe is controlled by the Einstein equations,
$G_{\mu \nu}=T_{\mu \nu}$. Since we expand the metric tensor in terms
of the perturbations, one must do the same for the Einstein tensor,
$G_{\mu \nu}=G_{\mu \nu}^{\rm FLRW}+\delta G_{\mu \nu}$ and for the
stress energy tensor,
$T_{\mu \nu}=T_{\mu \nu}^{\rm FLRW}+\delta T_{\mu \nu}$. Then, the
equations describing the behavior of the perturbations are
\begin{equation}
\delta G_{\mu \nu}=\delta T_{\mu \nu}.
\end{equation}
Of course, these equations are now partial differential equations
since the perturbations are supposed to describe the early
inhomogeneous and anisotropic Universe. But since these equations are
linear, they can be solved by going to Fourier space.

Then, the idea is to quantize the system. The motivation is that this
will provide a source for the cosmological perturbations (in other
words, this will fix the initial conditions). This source will be the
unavoidable quantum fluctuations of the inflaton and gravitational
fields at the beginning of inflation. On the technical front, this
means that $\delta g_{\mu \nu}$ will be promoted to a quantum
operator, $\delta g_{\mu \nu}\rightarrow \delta \hat{g}_{\mu \nu}$. As
consequence, curvature perturbations and gravitational waves also
become quantum operators, $\hat{\zeta}$ and $\hat{h}_{ij}$.

One fundamental assumption of inflation is that, initially, the
quantum perturbations are placed in the vacuum state. Then, this state
will evolve as the Universe expands. At the end of inflation, the
system will be placed into a strongly two-mode squeezed state. This
state is a very peculiar state and is defined as follows (here, we
follow the presentation of Ref.~\cite{Lvovsky:2014sxa}). Let us
consider a one-dimensional quantum oscillator. As is well-known, its
vacuum state is a Gaussian state whose wavefunction is given by
\begin{equation}
\Psi_0(x)=\frac{1}{\pi^{1/4}}e^{-x^2/2},
\end{equation}
where $x$ is the position of the oscillator. This state, written in
the momentum basis, reads
\begin{equation}
\tilde{\Psi}_0(p)=\frac{1}{\pi^{1/4}}e^{-p^2/2},
\end{equation}
where $p$ is the conjugate momentum of $x$. An interesting feature of
the vacuum state is that the dispersion in position and momentum are
equal, namely
\begin{equation}
\langle \Delta \hat{x}^2\rangle
=\langle \Delta \hat{p}^2\rangle=\frac12
\end{equation}
and saturates the Heisenberg inequality
$\langle \Delta \hat{x}^2\rangle \langle \Delta
\hat{p}^2\rangle=\frac14$.
A one-mode squeezed state is a also a Gaussian state but, in position
basis and momentum basis, its wave function is given by
\begin{equation}
\Psi_R(p)=\frac{\sqrt{R}}{\pi^{1/4}}e^{-R^2x^2/2},
\quad
\tilde{\Psi}_R(p)=\frac{1}{\pi^{1/4}\sqrt{R}}e^{-p^2/(2R^2)}.
\end{equation}
We see that the wavefunction now depends on an additional parameter,
$R$. As a consequence, the dispersion in position and momentum are no
longer equal,
\begin{equation}
\langle \Delta \hat{x}^2\rangle
=\frac{1}{2R^2}, \quad \langle \Delta \hat{p}^2\rangle=\frac{R^2}{2}
\end{equation}
although they still saturates the Heisenberg inequality. If $R>1$,
then the dispersion in position is smaller than that of the vacuum. We
say that the state is squeezed in position, hence its name. Of course,
since one has to satisfy the Heisenberg inequality, the price to pay
is that the dispersion on momentum is larger. If $R<1$, we have the
opposite situation and the state is squeezed in momentum.

Then, let us consider two oscillators. The vacuum state of this system
in position basis (namely the position of the first oscillator also
referred to as the position of Alice and the position of the second
oscillator also referred as to the position of Bob) can be written as
\begin{equation}
\Psi_{0}(x_1,x_2)=\frac{1}{\sqrt{\pi}}e^{-x_1^2/2-x_2^2/2}
=\frac{1}{\sqrt{\pi}}e^{-(x_1-x_2)^2/4}e^{-(x_1+x_2)^2/4}.
\end{equation}
We see that the position of Alice and Bob are uncorrelated. From this
expression, we are now in a position to introduce the two-mode
squeezed state which is given by
\begin{equation}
\Psi_{R}(x_1,x_2)=\frac{1}{\sqrt{\pi}}e^{-R^2(x_1-x_2)^2/4}e^{-(x_1+x_2)^2/(4R^2)},
\end{equation}
where the squeezing factor $R$ appears again and is related to the
squeezing parameter $r$ by $R=\ln r$. We see that the position of
Alice and Bob are now correlated. It is also interesting to notice
that the two-mode squeezed state does not imply squeezing for Alice or
Bob. Indeed, it is easy to check that
\begin{equation}
\langle \Delta \hat{x}_1^2\rangle
=\langle \Delta \hat{x}_2^2\rangle
=\frac{1+R^4}{4R^2}.
\end{equation}
These dispersions are always larger than those one would obtain from 
the vacuum state. This is related to the fact that, if one traces out, say, 
Alice's degree of freedom, the obtained state of Bob is not a 
one-mode squeezed state but a thermal state.

The quantum-mechanical properties of inflation discussed above are
clearly fascinating. Based on this aspect of the theory, one can
wonder whether it would be possible to exhibit quantum effects in the
sky. This was first discussed in
Refs.~\cite{Grishchuk:1990bj,Grishchuk:1992tw} and, more recently, in
Refs.~\cite{Martin:2012pea,Martin:2015qta,Martin:2016tbd,Martin:2016nrr,Martin:2017zxs,Martin:2018zbe,Martin:2018lin}.

Let us now turn to a quantitative characterization of the cosmological
fluctuations originating from inflation. As usual this will be done by
computing the various correlation functions of scalar and tensor
perturbations (in the following, we mainly focus on the scalar
sector). The simplest correlation function is evidently the two-point
correlation function which is given by
\begin{equation}
\label{eq:meanzetasquare}
  \langle \zeta^2(\eta, {\bm x})\rangle =\int _0^{+\infty}
\frac{{\rm d}k}{k}
  \calP_{\zeta}(k),
  \end{equation}
where brackets mean quantum averages in the two mode squeezed state 
described above and where $\calP_{\zeta}(k)=k^3\vert \zeta_{\bm k}
\vert^2/(2\pi^2)$ is, by definition, the power spectrum of scalar 
perturbations. This scalar power spectrum is a very important 
quantity because it can be probed observationally by measuring the 
CMB anisotropies or by measuring the distributions of galaxies across 
our Universe. Using the slow-roll approximation introduced above, 
it can also be calculated for an arbitrary potential $V(\phi)$ and 
the result reads
\begin{equation} 
\label{eq:spectrumsr}
\calP_\zeta(k)= \calP_{\zeta 0}(k_{_{\rm P}})
\left[a_0^{_{({\rm S})}} + 
a_1^{_{({\rm S})}} \ln \left(\dfrac{k}{k_{_{\rm P}}}\right) 
+ \frac{a_2^{_{({\rm S})}}}{2} \ln^2\left(\dfrac{k}{k_{_{\rm P}}}\right)
+ \cdots \right]\, ,
\end{equation}
where $k_{_{\rm P}}$ is a pivot scale and the
global amplitude can be expressed as
\begin{equation}
\label{eq:scalaramp}
\calP_{\zeta \zero} =\frac{H_*^2}{8 \pi^2 \epsilon_{1*} \Mp^2}\,.
\end{equation}
In the above formula, a star means that the corresponding quantity has
been calculated at the time at which the pivot scale crossed out the
Hubble radius during inflation.  We notice that the amplitude of the
correlation function depends on the square of the Hubble rate during
inflation (measured in Planck units) and is inversely proportional to
the first slow-roll parameter. All these quantities are scale
independent and so is the global amplitude.  This result is viewed as
one of the most important success of inflation.  Indeed, before the
invention of inflation, it was already known that a scale invariant
power spectrum (or Harrisson-Zeldovitch power spectrum) is a good fit
to the data. But its origin was mysterious and there was no convincing
physical mechanism to produce it. Inflation, on the contrary,
naturally implies this property.  In fact, generically, exact scale
invariance is not a prediction of inflation because, as can be seen in
Eq.~(\ref{eq:spectrumsr}), the overall amplitude receives small, scale
dependent, logarithmic corrections. The amplitudes of those
corrections is determined by the Hubble flow parameters,
namely~\cite{Schwarz:2001vv,Casadio:2004ru,Casadio:2005xv,Casadio:2005em,Gong:2001he,Choe:2004zg,Leach:2002ar,Lorenz:2008et,Martin:2013uma},
\begin{eqnarray}
\label{eqn:as0}
a_{0}^{\usssPSP} &=& 1 - 2\left(C + 1\right)\epsilon_{1*} - C \epsilon_{2*}
+ \left(2C^2 + 2C + \frac{\pi^2}{2} - 5\right) \epsilon_{1*}^2 \nonumber
\\ & + & \left(C^2 - C + \frac{7\pi^2}{12} - 7\right)
\epsilon_{1*}\epsilon_{2*} + \left(\frac12 C^2 + \frac{\pi^2}{8} -
1\right)\epsilon_{2*}^2 \nonumber \\ & + & \left(-\frac12 C^2  +
\frac{\pi^2}{24}\right)  \epsilon_{2*}\epsilon_{3*} +\cdots \, , \\
\label{eqn:as1}
a_{1}^{\usssPSP} & = & - 2\epsilon_{1*} - \epsilon_{2*} 
+ 2(2C+1)\epsilon_{1*}^2
+ (2C - 1)\epsilon_{1*}\epsilon_{2*} + C\epsilon_{2*}^2 
- C\epsilon_{2*}\epsilon_{3*}
+\cdots \, ,\\ 
a_{2}^{\usssPSP} &=& 4\epsilon_{1*}^2 + 2\epsilon_{1*}\epsilon_{2*} +
\epsilon_{2*}^2 - \epsilon_{2*}\epsilon_{3*} +\cdots \, ,\\
a_{3}^{\usssPSP} &=& \calO(\epsilon_{n*}^3)\, ,
\label{eqn:as2}
\end{eqnarray}
where $C \equiv \gamma_{\usssE} + \ln 2 - 2 \approx -0.7296$,
$\gamma _\usssE$ being the Euler constant. Therefore, the exact prediction 
of inflation (really a prediction since it was made before it was checked) 
is that the power spectrum should be almost scale 
invariant but not exactly scale invariant. This prediction has been 
recently confirmed for the first time by the Planck data. Technically, one 
defines the spectral index, which is the logarithmic derivative
of $\ln \calP_{\zeta}(k)$, namely
\begin{equation}
\label{eq:specindices}
\nS=1-2\epsilon_{1*}-\epsilon_{2*},
\end{equation}
where $\nS=1$ corresponds to exact scale invariance. As will be discussed 
in more details in the following, Planck has measured $\nS\simeq 0.96$ and 
$\nS=1$ is now excluded at more than $5\sigma$. We also see that 
the spectral index depends on the two first Hubble flow parameters. As a 
consequence, a measurement of $\nS$ is also a measurements of $\epsilon_{1*}$ 
and $\epsilon_{2*}$, that is to say of the first and second derivative 
of the inflaton potential. This explains how astrophysical measurements 
can constrain the theory of inflation.

The treatment of tensor modes (primordial gravitational waves) proceeds 
in the very same way. One can compute the two-point correlation and the 
power spectrum using the slow-roll approximation. One then arrives at 
the following expression
\begin{equation} 
\label{spectrumsr}
\calP_h(k)= \calP_{h 0}(k_{_{\rm P}})
\left[a_0^{_{({\rm T})}} + 
a_1^{_{({\rm T})}} \ln \left(\dfrac{k}{k_{_{\rm P}}}\right) 
+ \frac{a_2^{_{({\rm T})}}}{2} \ln^2\left(\dfrac{k}{k_{_{\rm P}}}\right)
+ \cdots \right]\, ,
\end{equation}
where the amplitude $\calP_{h 0}(k_{_{\rm P}})$ is given by
\begin{equation}
\label{eq:tensoramp}
\calP_{h \zero} =\frac{2 H_*^2}{\pi^2 \Mp^2}\,.
\end{equation}
As it was the case for scalar perturbations, the overall amplitude is
also given by the square of the expansion rate during inflation
measured in Planck units. Of course the big difference is that the
first slow-roll parameter $\epsilon_{1*}$ is now absent. This means
that a measurement of the tensor modes would immediately provide the
energy scale of inflation.  Notice that $\calP_{h 0}(k_{_{\rm P}})$ is
also scale independent and, at leading order, the tensor power
spectrum is therefore scale invariant.  However, as it was also the
case for scalar modes, this scale invariant amplitude receives small,
scale dependent, logarithmic corrections the amplitude of which can be
expressed as~\cite{Leach:2002ar}
\begin{eqnarray}
a_{0}^{\usssPTP} &=&
 1 - 2\left(C + 1\right)\epsilon_{1*} 
 + \left(2C^2 + 2C + \frac{\pi^2}{2} - 5\right) 
 \epsilon_{1*}^2 \nonumber \\ 
& + & \left(-C^2 - 2C + \frac{\pi^2}{12} - 2\right) 
 \epsilon_{1*}\epsilon_{2*} +\cdots \, ,\\
a_{1}^{\usssPTP} &=& 
 - 2\epsilon_{1*} + 2(2C + 1)\epsilon_{1*}^2 
 - 2(C + 1)\epsilon_{1*}\epsilon_{2*}+\cdots  \, ,
\label{eqn:at1} \\
a_{2}^{\usssPTP} &=& 4\epsilon_{1*}^2 - 2\epsilon_{1*}\epsilon_{2*}
+\cdots  \, , \\
a_{3}^{\usssPTP} &=& \calO(\epsilon_{n*}^3)\, .
\label{eqn:at2}
\end{eqnarray}
From the coefficient $a_{1}^{\usssPTP}$, one can read the tensor spectral 
index (at first order in slow-roll). One obtains
\begin{equation}
\label{eq:gwspecindices}
\nT=-2\epsilon_1.
\end{equation}
Exact scale invariance corresponds to $\nT=0$ (for historical reasons, the 
convention differs from that of scalars). Another difference is that $\nT$ 
depends on $\epsilon_{1*}$ only while $\nS$ depends on $\epsilon_{1*}$ and 
$\epsilon_{2*}$. Given that $\epsilon_{1*}$ is always positive, this 
implies that $\nT$ is always negative (or red).

Finally, one can also calculate the tensor amplitude to scalar 
amplitude $r$. Using the previous expressions, one obtains
\begin{equation}
\label{eq:defr}
r \equiv \frac{\calP_h}{\calP_\zeta}=16\epsilon_{1*}.
\end{equation}
Since, by definition, $\epsilon_{1*}\ll 1$, this means that 
gravitational waves are sub-dominant (which explains why they have 
not yet been detected~\cite{Ade:2015tva,Martin:2014lra}). Notice that 
there is a priori no lower bound on $r$. Therefore, if $r$ turns out 
to be very small, primordial gravitational waves will probably never 
been detected but this would be in no way in contradiction with the 
predictions of inflation. At the time of writing, it is believed 
that the next generations of telescope and satellites will be able 
to reach the level $r\sim 10^{-3}$ maybe a bit smaller. Let us 
hope that Nature has produced a $r$ larger than this limit!

To conclude this section, let us mention Non-Gaussianities (NG). So
far, we have restricted our considerations to two-point correlation
functions. Of course, higher correlation functions are also of great
interest. Usually, the three-point correlation function (bispectrum)
and the four-point correlation function (trispectrum) are considered.
For the models described previously, NG are very small (of the order
of the slow-roll
parameters)~\cite{Gangui:1993tt,Gangui:1994yr,Gangui:1999vg,
  Maldacena:2002vr}. The reason is easy to understand. We have started
from a Gaussian state and the evolution of the perturbations is
linear. As a consequence, the appearance of any NG is necessarily
related to non linearities, which are very small.

\section{Extensions}
\label{sec:extensions}

So far, we have described the simplest way to realize
inflation. However, since the invention of inflation in the $80$'s,
more complicated scenarios have been imagined. In this section, we say
a few words about them.

The most generic extension is probably to consider models where,
instead of having one scalar field, one has several ones playing an
active role during inflation~\cite{Wands:2007bd}. This appears to be a
natural approach given that inflation can occur at energy scales as
high as $10^{15}\GeV$. At those scales, it is believed that particle
physics is no longer described by the standard model but by its
extensions (SUSY, SUGRA, string theory, etc \dots). And, usually, in
these alternative frameworks, there are plethora of scalar fields.

Clearly, multiple field inflation scenarios are more complicated and
it is more difficult to make generic predictions. However, one can
list three main modifications. Firstly, there is the possibility of
having non adiabatic perturbations, which is impossible for single
field models. The reason is that, if several scalar fields are present
during inflation, then the corresponding decay products can have
different origin resulting in the possible presence of non adiabatic
perturbations. Secondly, non adiabatic perturbations can source the
evolution of curvature perturbations. As a result, if they are present
during inflation and reheating, $\zeta(\eta,{\bm x})$ on large scales
is no longer a conserved quantity. This has drastic consequences,
especially for reheating, which then becomes potentially dependent on
the details of physical processes going on on scales smaller than the
Hubble radius. Thirdly, it is possible to produce non negligible
NG. As already mentioned, these modifications are not mandatory and
must be analyzed on a model by model basis.

Yet other extensions are also possible such that having a non
canonical kinetic term for the scalar field. They are called
K-inflation models~\cite{ArmendarizPicon:1999rj,Garriga:1999vw} (for
the observational status of this class of models, see
Refs.~\cite{Lorenz:2008et,Lorenz:2008je,Martin:2013uma}). It is also
possible to have models with
features~\cite{Starobinsky:1992ts,Hazra:2010ve}. This means a model of
inflation where, in some limited region, the potential is not
flat. This usually causes a transitory violation of the slow-roll
approximation which can result in oscillations in the power spectrum
and non negligible
NG~\cite{Martin:2011sn,Hazra:2012yn,Martin:2014kja}.  More complicated
models are possible, for instance by combining the various ingredients
discussed above~\cite{Avila:2013ela}, but we will not discuss them
here. We now turn to another question, namely how the observations can
discriminate among these various possibilities.

\section{Inflation and CMB Observations}
\label{sec:observations}

\begin{figure}
\begin{center}
\includegraphics[width=15cm]{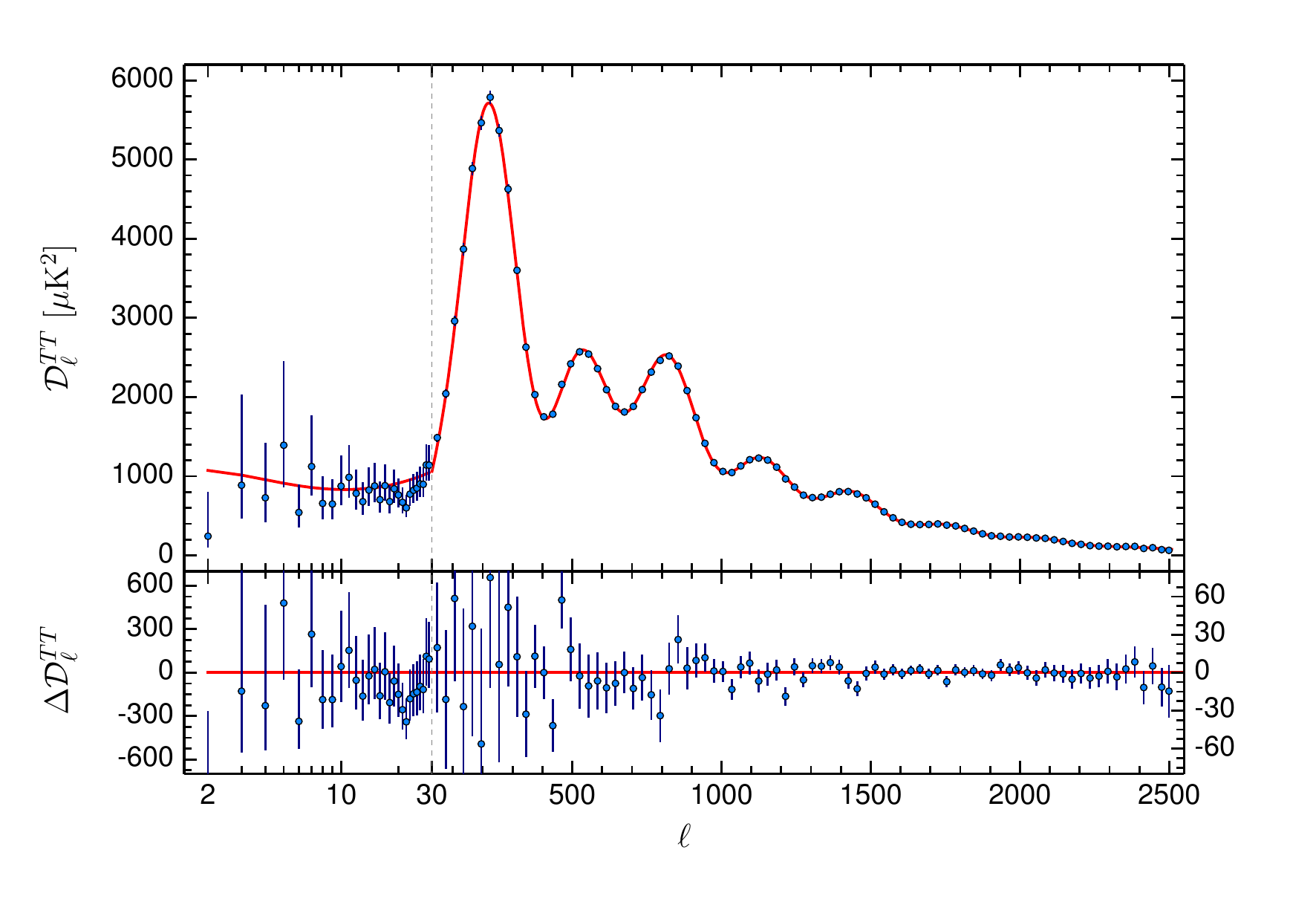}
\end{center}
\caption{Multipole moments versus angular scale from Planck $2015$
  data. The multipole moments are obtained from the CMB map by Fourier
  transforming it according to:
  $\langle \delta T/T({\bm e}_1)\delta T/T({\bm
    e}_2)\rangle=(4\pi)^{-1}\sum _{\ell}(2\ell+1)C_{\ell}P_{\ell}(\cos
  \theta)$
  where $\theta$ is the angle between two directions ${\bm e}_1$ and
  ${\bm e}_2$ and $P_\ell$ is a Legendre polynomial. The multipole
  moments $C_{\ell}$ are interpreted as the power of the signal at a
  given angle $\theta$. Notice that ${\cal D}_{\ell}$ is related to
  $C_\ell$ by ${\cal D}_{\ell}=\ell(\ell+1)C_{\ell}/(2\pi)$. The red
  curve corresponds to the best fit and is consistent with the
  predictions of single field, slow-roll, inflation. Figure taken from
  Ref.~\cite{Ade:2015xua}.}
\label{fig:TTplanck2015}
\end{figure}

\begin{figure}
\begin{center}
\includegraphics[width=13cm]{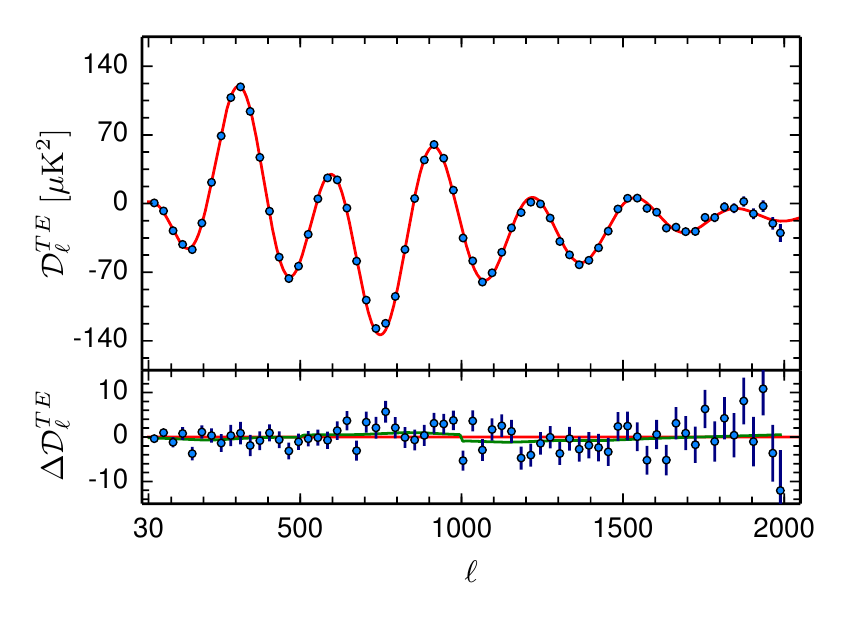}
\end{center}
\caption{Multipole moments corresponding to the correlation between
  temperature and $E$-mode polarization anisotropies. The red solid
  line is obtained from temperature measurements only, see
  Fig.~\ref{fig:TTplanck2015}. The lower panel shows the residual with
  respect to this best fit. Figure taken from
  Ref.~\cite{Ade:2015xua}.}
\label{fig:TEplanck2015}
\end{figure}

\begin{figure}
\begin{center}
\includegraphics[width=13cm]{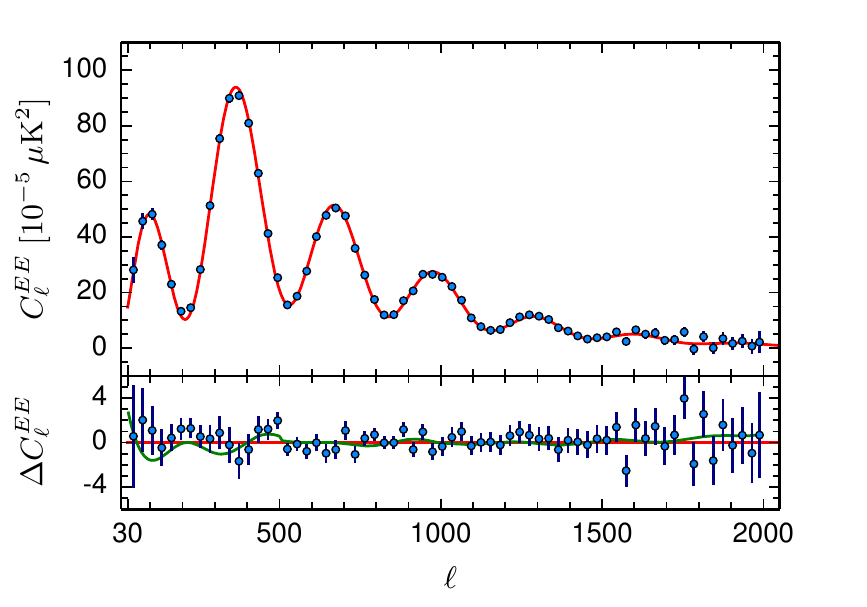}
\end{center}
\caption{Same as in Fig.~\ref{fig:TEplanck2015} but for the $E$-mode
  power spectrum obtained from Planck $2015$. Figure taken from
  Ref.~\cite{Ade:2015xua}.}
\label{fig:EEplanck2015}
\end{figure}

The Planck satellite has recently measured the CMB temperature, see
Fig.~\ref{fig:TTplanck2015}, and polarization, see
Figs.~\ref{fig:TEplanck2015} and~\ref{fig:EEplanck2015}, anisotropies
with unprecedented accuracy. These new data allow us to constrain
inflation and to learn which was version of inflation realized in the
early Universe.

In brief, Planck has shown that the Universe is spatially flat, that
the perturbations are adiabatic and Gaussian~\cite{Ade:2015lrj}. These
results are all consistent with single field (with minimal kinetic
term), slow-roll, inflation which, therefore, appears to be the
preferred class of models. This does not mean that the more
complicated versions discussed in Section~\ref{sec:extensions} are
ruled out but just that, at the moment, they are not needed in order
to explain the data.

With regards to inflation, probably the most important discovery made
by the Planck satellite is the measurement of the scalar spectral
index~\cite{Ade:2015lrj}
\begin{equation}
\nS=0.969\pm 0.005.
\end{equation}
For the first time, the value $\nS=1$ is excluded at more than
$5\sigma$. As was already discussed above, the fact that the power
spectrum must be scale invariant (the so called Harrisson-Zeldovitch
power spectrum) was known long ago (before the invention of
inflation). But the non trivial prediction of inflation was that $\nS$
should be close to one but not exactly one. And this is exactly what
has been observed for the first time by the Planck satellite.

Another important of piece of information is that, unfortunately, so
far, no gravitational waves has been detected. This means the
following upper bound on the tensor to scalar ratio
$r$~\cite{Ade:2015tva}
\begin{equation}
r\lesssim 0.08.
\end{equation}

\begin{figure}
\begin{center}
\includegraphics[width=13cm]{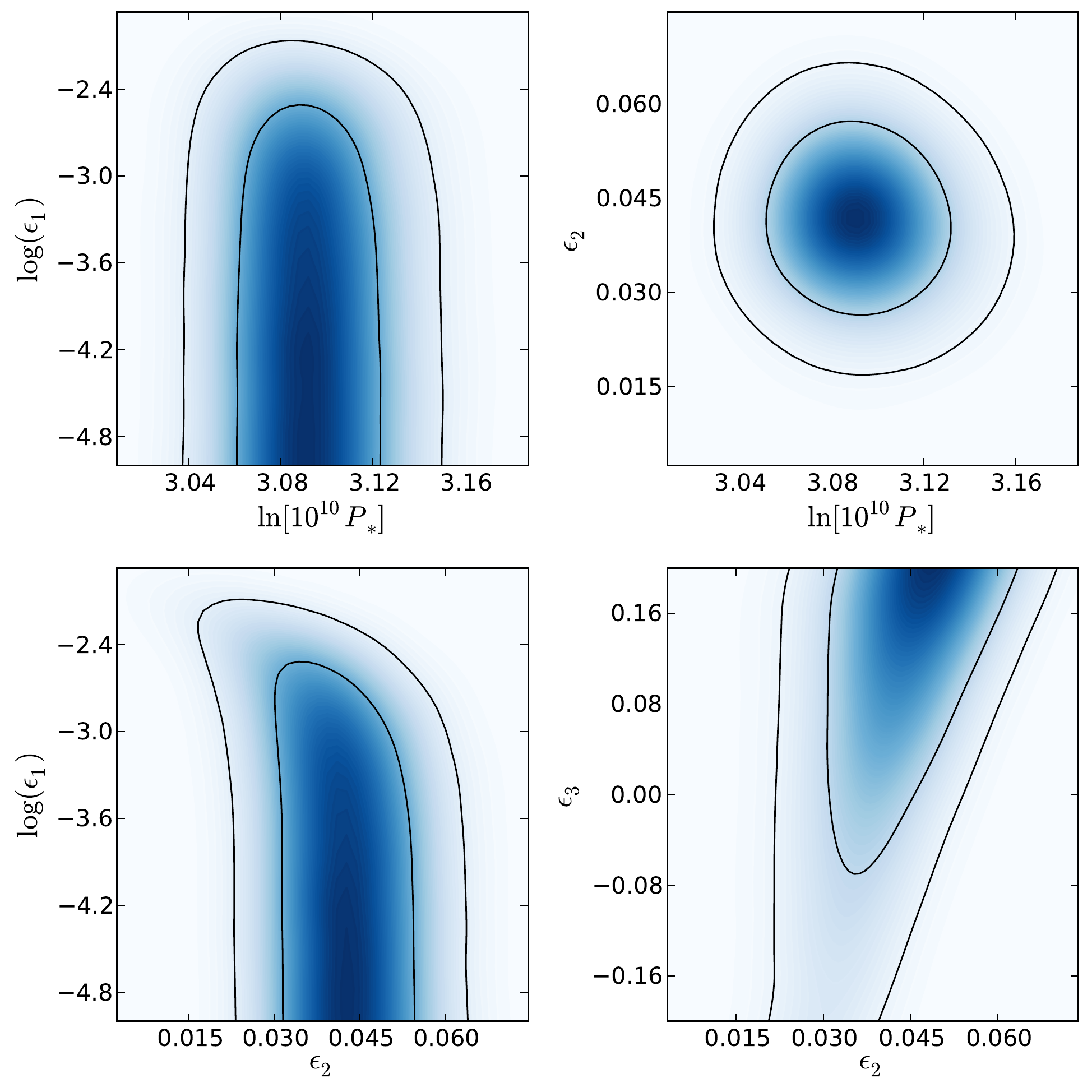}
\end{center}
\caption{Posterior distributions of the parameters $\epsilon_{1*}$,
  $\epsilon_{2*}$, $\epsilon_{3*}$ and
  $P_*\equiv a_0^{_{({\rm S})}}\calP_{\zeta 0}$. The posteriors are
  taken to be Jeffreys's priors for $P_*$ and $\epsilon_{1*}$ and flat
  priors for $\epsilon_{2*}$ and $\epsilon_{3*}$.}
\label{fig:sr}
\end{figure}

From the measurements of those quantities, one can also infer
constraints on the Hubble flow parameters, see Fig.~\ref{fig:sr} and
Refs.~\cite{Martin:2013tda,Martin:2013nzq,Martin:2015dha}. We see that
$P_*\equiv \calP _{\zeta 0}a_0^{_{\rm (S)}}$ and $\epsilon_{2*}$ are
constrained while there only exists an upper bound on
$\epsilon_{1*}$. Of course, $P_*$ is determined because one knows the
amplitude of CMB fluctuations (namely $\delta T/T\simeq 10^{-5}$).  On
the other hand, the upper bound on $\epsilon_{1*}$ originates from
Eq.~(\ref{eq:defr}) and the fact that we only have an upper bound on
$r$.  Given that $H_*^2/\Mp^2\simeq 8\pi^2\epsilon_{1*}P_*$, this
means that we only have an upper bound on the energy scale of
inflation, namely
\begin{equation}
H_* \lesssim 1.2\times 10^{14}\GeV,
\end{equation}
or $\rho_*^{1/4}\lesssim 2.2\times 10^{16}\GeV$. Finally, the third
slow-roll parameter, $\epsilon_{3*}$ is not well constrained which
means that we do not have yet a detection of a running.

\begin{figure}
\begin{center}
\includegraphics[width=6cm]{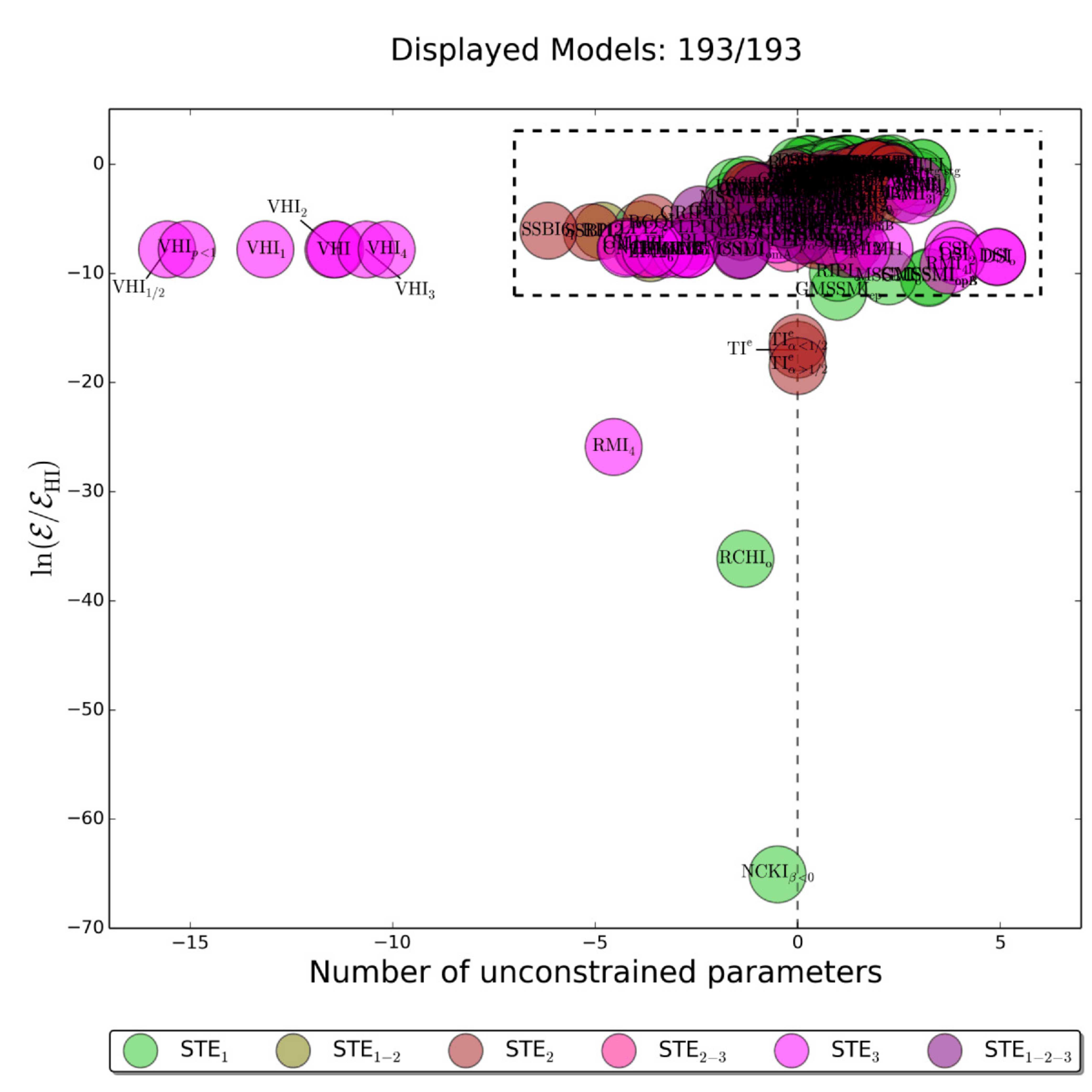}
\includegraphics[width=6cm]{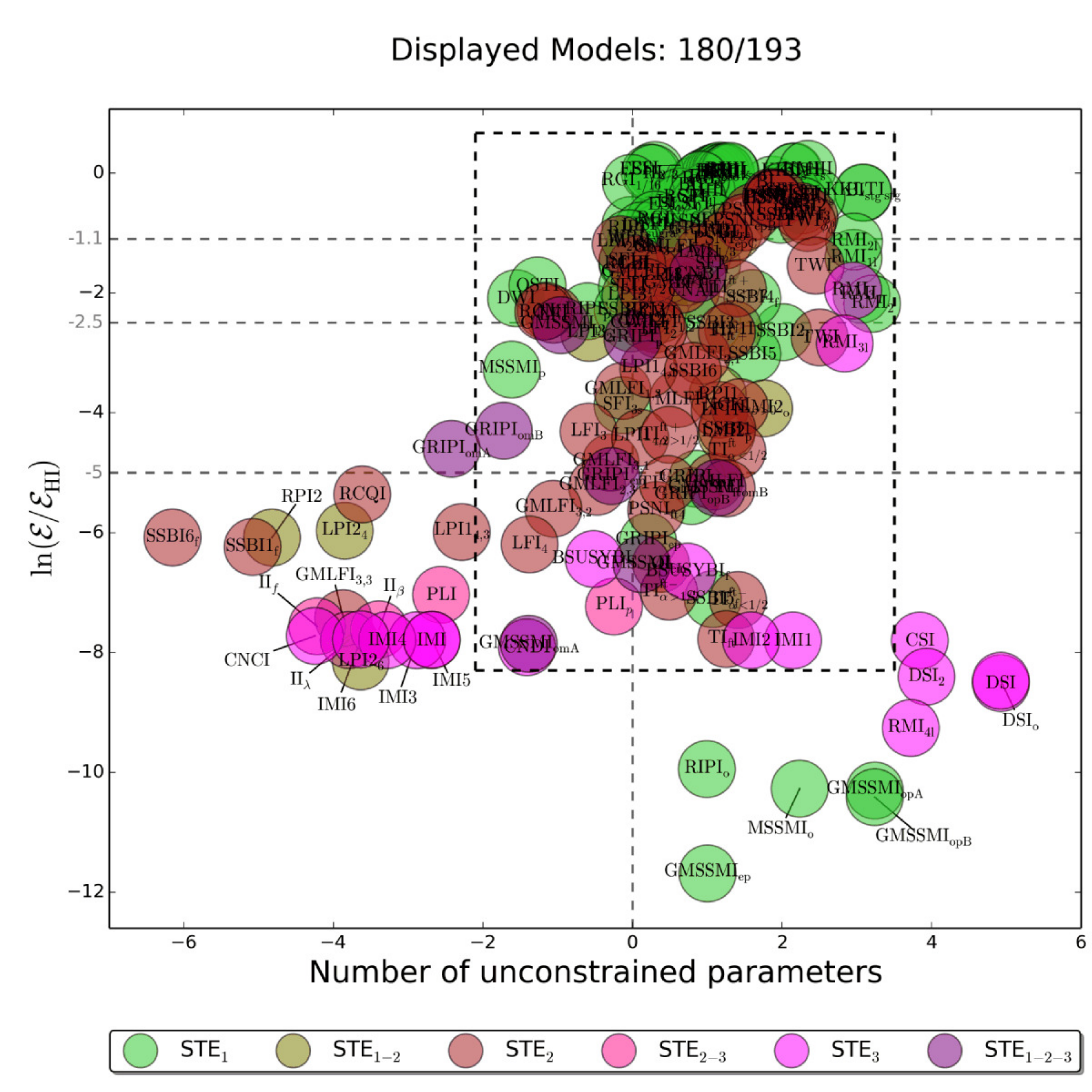}
\includegraphics[width=6cm]{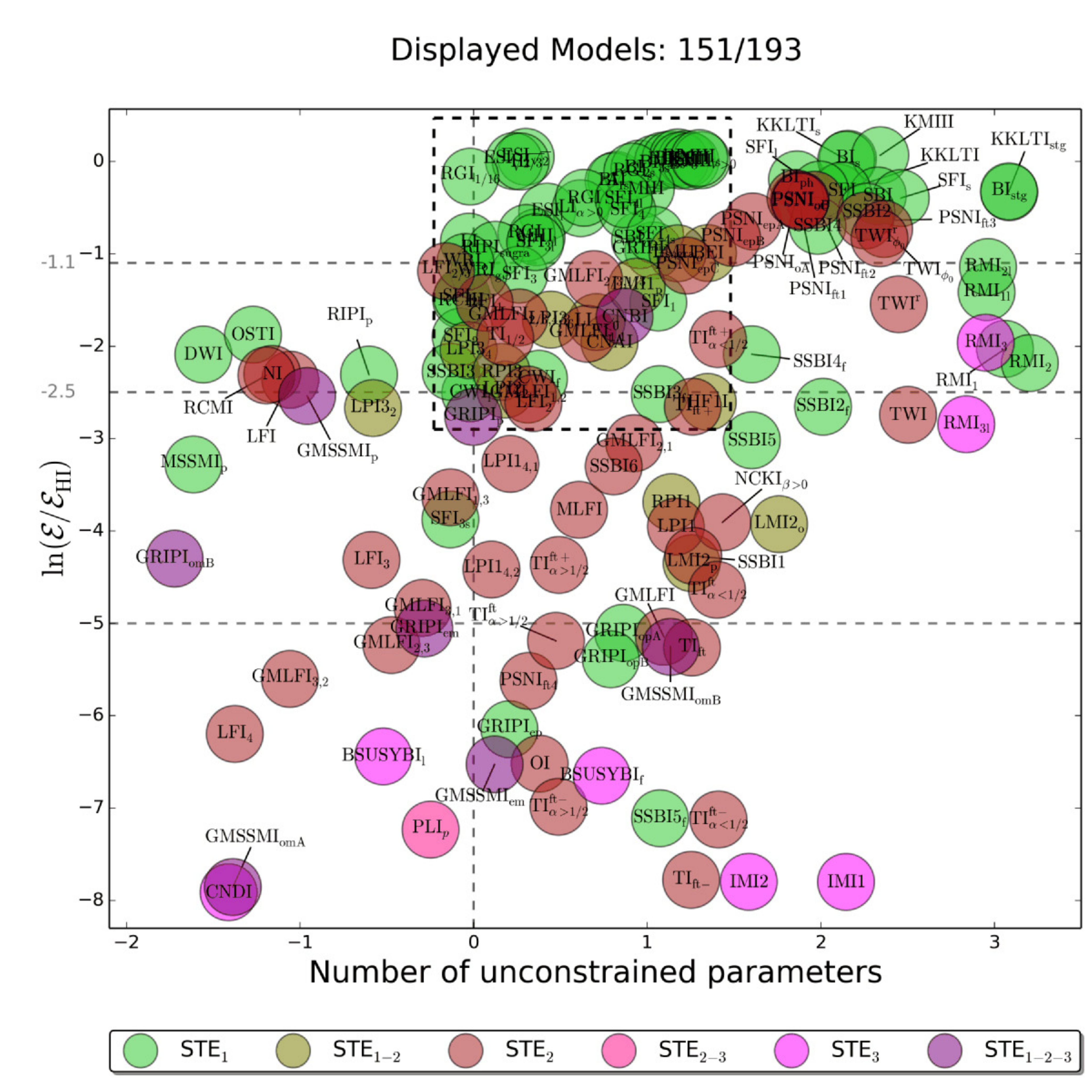}
\includegraphics[width=6cm]{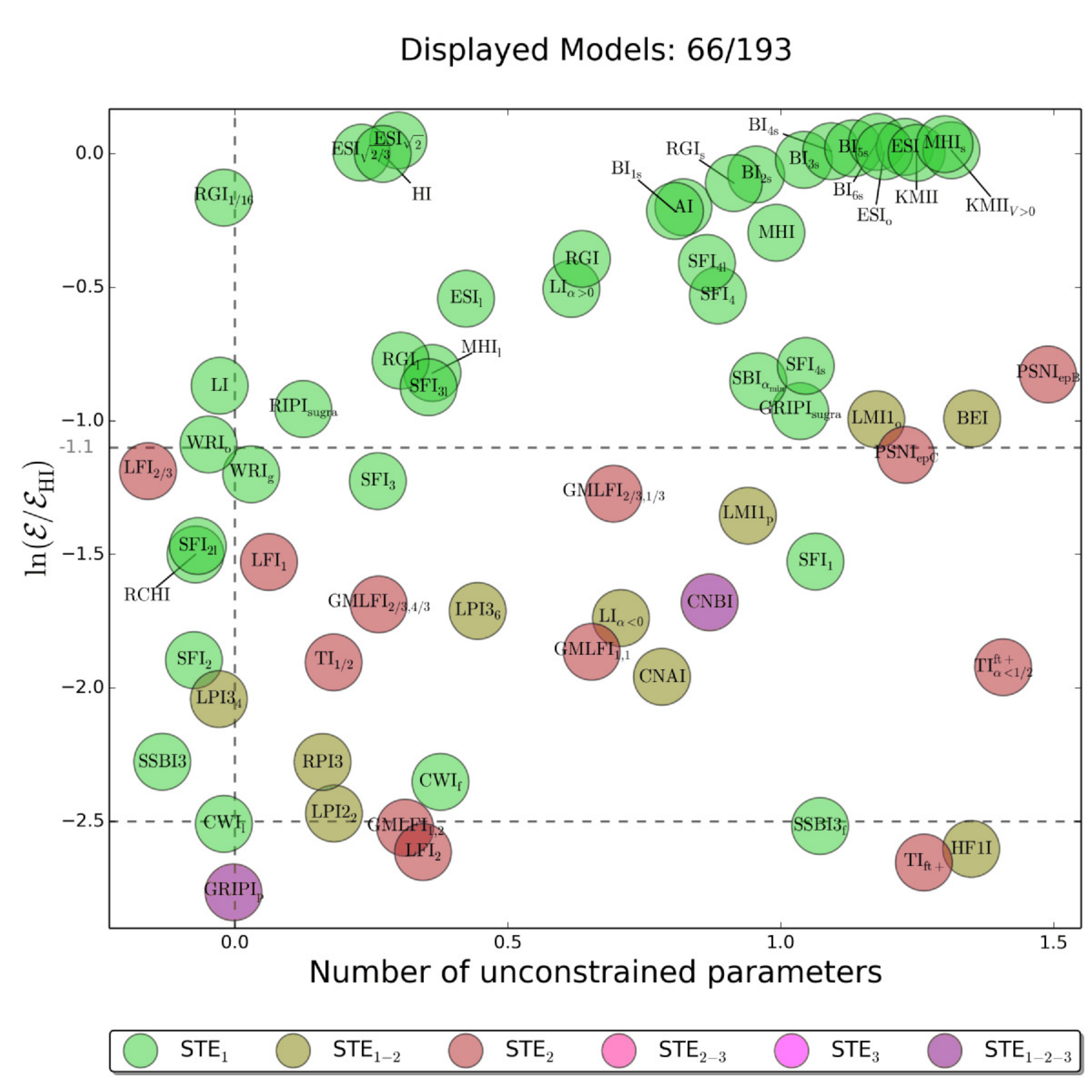}
\end{center}
\caption{Bayesian evidence versus number of unconstrained parameters
  for different models of inflation. Each circle represents a given
  inflationary scenario (the size of the circle has no meaning). The
  upper right panel is a zoom on the ``best'' region (the square
  delimited by the dashed black line) of the upper left panel. In the
  same way, the bottom left plot is a zoom on the ``best'' region of
  the upper right. Finally, the bottom right is a zoom on the ``best''
  region of the bottom left figure.}
\label{fig:evidence}
\end{figure}

We have seen before that the slow-roll parameters carry information
about the shape of the inflaton potential. Since we have obtained
constraints on these parameters, we must be able to say something about
the shape of the inflaton potential
itself~\cite{Martin:2013tda,Martin:2013nzq,Martin:2015dha}. In order
to answer this question, one can calculate the Bayesian evidence of
the various models of inflation. The Bayesian evidence is the integral
of the likelihood function over the prior space. It characterizes the
performance of a model and its ability to fit the
data~\cite{Trotta:2008qt}. The larger the evidence, the better the
model. In Refs.~\cite{Martin:2013tda,Martin:2013nzq,Martin:2015dha},
the Bayesian evidence of nearly two hundred models were computed. The
result of this computation is displayed in Figs.~\ref{fig:evidence}
where the number of unconstrained parameters is also indicated. A
detailed analysis of those results has been published in
Refs.~\cite{Martin:2013tda,Martin:2013nzq,Martin:2015dha}, but the
bottom line is that plateau inflationary models are the ``best''
models according to the Planck data. A plateau potential is a
potential which flattens out at infinity. The prototype of this class
of models is the so-called Starobinsky model given by
\begin{equation}
V(\phi)=M^4\left(1-e^{-\sqrt{2/3}\phi/\Mp}\right)^2.
\end{equation}
This conclusion is non trivial since models that were historically
considered as leading candidates, such as $V(\phi)=m^2\phi^2/2$, are
now strongly disfavored compared to plateau models.

Let us also notice another interesting point. The prediction of
plateau models for $r$ is, roughly speaking, $r\simeq 10^{-3}$. As
indicated before, this value is in principle reachable by the next
generation of instruments. This means that there is maybe a good
chance to detect primordial gravitational in a non too distance future
(say, a decade).

\begin{figure}
\begin{center}
\includegraphics[width=13cm]{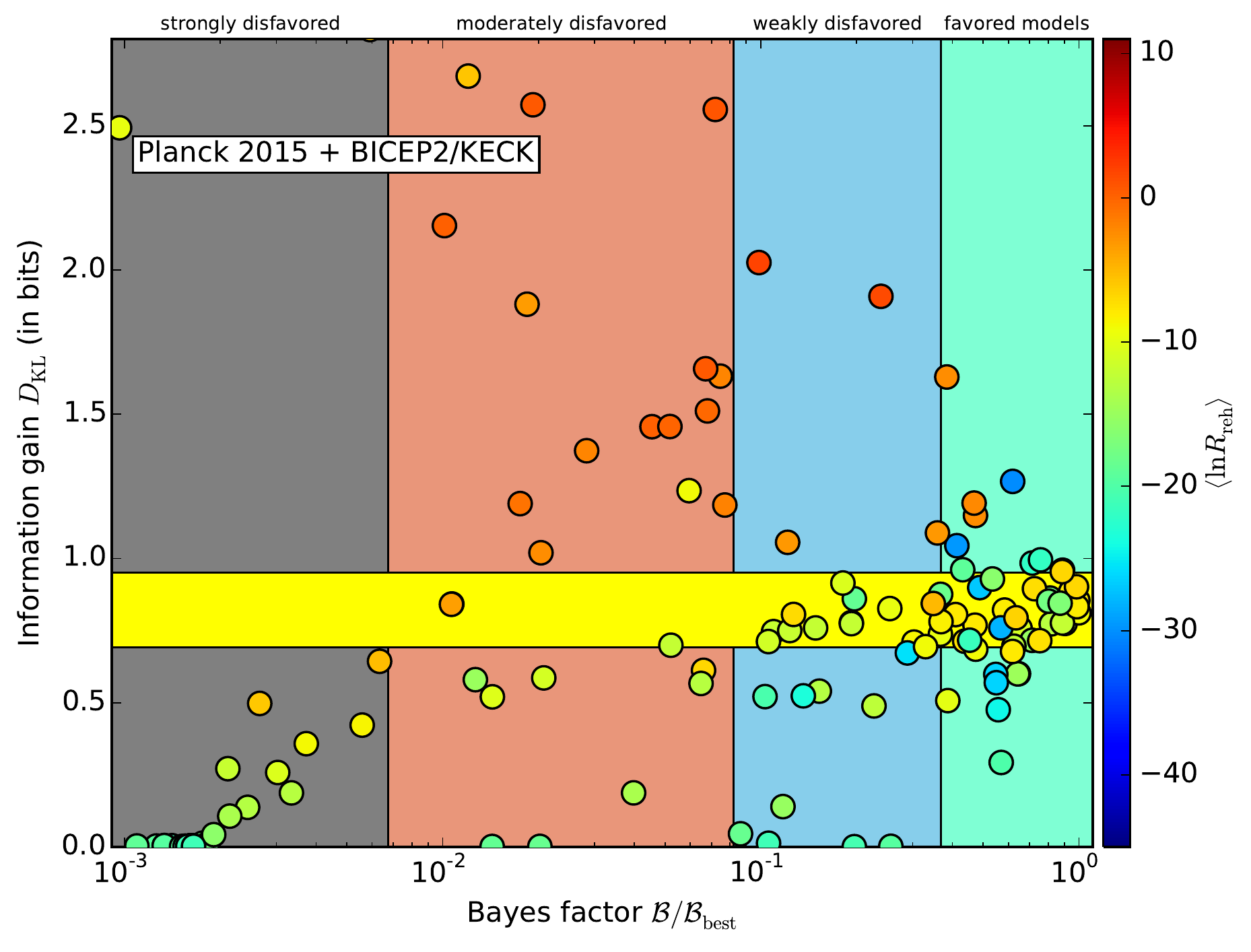}
\end{center}
\caption{Kullback-Leibler $D_{_{\rm KL}}$ divergence versus Bayesian
  evidence for various models of inflation. The mean value is given 
by $\langle D_{_{\rm KL}}\rangle =0.82\pm 0.13$ and the yellow band 
represents the one-sigma deviation around this mean value.}
\label{fig:reheating}
\end{figure}

Finally, let us discuss what the Planck data imply for reheating. As
was discussed before, constraints on reheating are expressed through
constraints on the reheating parameter $\Rrad$ defined in
Eq.~(\ref{eq:defRrad}). In
Refs.~\cite{Martin:2006rs,Martin:2010kz,Martin:2014nya,Martin:2016oyk},
the posterior distributions was derived for the nearly two hundred
models already considered before for the calculation of the Bayesian
evidence. The situation is summarized in Fig.~\ref{fig:reheating}. It
represents the Kullback-Leibler divergence between the prior
distribution and the posterior versus the Bayesian evidence for
different models of inflation (represented by circles). The
Kullback-Leibler divergence is defined by
\begin{equation}
D_{_{\rm KL}}=\int P\left(\ln R_{\rm reh}\vert D\right)
\ln \left[\frac{P\left(\ln R_{\rm reh}\vert D\right)}
{\pi\left(\ln R_{\rm reh}\right)}\right]{\rm d}\ln R_{\rm reh},
\end{equation}
where $R_{\rm reh}$ is given by
$\ln R_{\rm reh}=\ln R_{\rm rad}+\ln (\rho_{\rm end}/\Mp^4)/4$ and is
therefore, for a given model of inflation, in a one-to-one
correspondence with $\Rrad$. The quantity $\pi$ represents the prior
on $R_{\rm reh}$ and $P$ the posterior. The Kullback-Leibler
divergence measures the ``distance'' between the prior and the
posterior and, as a consequence, also represents the amount of
information provided by the data $D$ (of course, here, the Planck
data) about $\ln R_{\rm reh}$. The constraints are model dependent and
one has a posterior distribution per model of inflation, an amount of
information which, given the number of scenarios analyzed, is
difficult to deal with. The value of $D_{_{\rm KL}}$ is one way to
summarize the information about reheating for a given model to one
number. In this sense, Fig.~\ref{fig:reheating} completely describes
what, for each known model of inflation, the Planck data implies with
regards to the ability to fit the data and to reheating. Let us also
notice that one can calculate the mean value of $D_{_{\rm KL}}$. One
finds $\langle D_{_{\rm KL}}\rangle=0.82\pm 0.13$, which expresses the
fact that reheating is globally constrained by the Planck data.

\section{Conclusions}
\label{sec:conclusions}

In this short review, we have discussed the theory of inflation. Over
the years, the inflationary scenario has become a crucial ingredient
in our understanding of Cosmology. It is important to stress that
inflation is not an alternative to the standard model of Cosmology, it
is rather a new part of it.

Invented in the $80$'s, inflation has recently witnessed new
developments with the publication of the high accuracy Planck
data. Clearly, these data have boosted our confidence in inflation. In
particular, the measurement of the spectral index to be close but not
equal to one is an important confirmation of an inflationary
prediction. Admittedly, it is probably not the final proof that
inflation actually occurred in the early Universe but it nevertheless
represents a very strong argument in its favor. From the Planck data,
we have also learned that inflation is probably realized in its
simplest version (single field, slow-roll, with minimal kinetic term)
and that the best scenario is a plateau model for which the potential
flattens out at very large values of the field.

What is then the next step? Clearly, the detection of primordial
gravitational waves will play a crucial role. It is an unambiguous
prediction of inflation that has not yet been confirmed. Future
missions will be able to reach $r\sim 10^{-3}$. Unfortunately,
inflation, as a paradigm, does not predict the value of $r$ even if
$r$ is predicted if a precise scenario is given. However, the best
model of inflation, the Starobinsky model, predicts a value of $r$
which, in principle, could be detected in the future. 

Let us also add that the detection of NG will also certainly play an
important role in the future. Given that we deal with the simplest
class of models, the expected signal is very small and its detection
will be challenging (if possible). But, obviously, this would be of
crucial importance.

Of course, inflation is not a perfect scenario and some of its aspects
remain unclear. But, as an effective model of the early Universe, it
scores pretty well. Let us see whether its performances remain so
efficient in the future.

\acknowledgments I would like to thank the organizers, especially
Profs. J.~Silk and N.~Vittorio, and the Italian Physical Society (SIF)
for having invited me to lecture at the International School of
Physics “Enrico Fermi” (from 26th June to 19th July 2017) at the
beautiful Villa Monastero located in Varenna, Lake of Como (Italy).

\bibliographystyle{varenna}
\bibliography{varenna_jmartin}

\begin{thebibliography}{10}
\expandafter\ifx\csname url\endcsname\relax\def\url#1{\texttt{#1}}\fi
\expandafter\ifx\csname urlprefix\endcsname\relax\def\urlprefix{URL }\fi

\bibitem{Starobinsky:1980te}
\NAME{Starobinsky A.~A.}, \IN{Phys. Lett.B}{91}{1980}{99}.
%%CITATION = PHLTA,B91,99;%%

\bibitem{Starobinsky:1982ee}
\NAME{Starobinsky A.~A.}, \IN{Phys. Lett.B}{117}{1982}{175}.
%%CITATION = PHLTA,B117,175;%%

\bibitem{Guth:1980zm}
\NAME{Guth A.~H.}, \IN{Phys. Rev.D}{23}{1981}{347}.
%%CITATION = PHRVA,D23,347;%%

\bibitem{Linde:1981mu}
\NAME{Linde A.~D.}, \IN{Phys. Lett.B}{108}{1982}{389}.
%%CITATION = PHLTA,B108,389;%%

\bibitem{Albrecht:1982wi}
\NAME{Albrecht A. \atque Steinhardt P.~J.}, \IN{Phys. Rev.
  Lett.}{48}{1982}{1220}.
%%CITATION = PRLTA,48,1220;%%

\bibitem{Linde:1983gd}
\NAME{Linde A.~D.}, \IN{Phys. Lett.B}{129}{1983}{177}.
%%CITATION = PHLTA,B129,177;%%

\bibitem{Starobinsky:1979ty}
\NAME{Starobinsky A.~A.}, \IN{JETP Lett.}{30}{1979}{682}.
%%CITATION = JTPLA,30,682;%%

\bibitem{Mukhanov:1981xt}
\NAME{Mukhanov V.~F. \atque Chibisov G.}, \IN{JETP Lett.}{33}{1981}{532}.

\bibitem{Mukhanov:1982nu}
\NAME{Mukhanov V.~F. \atque Chibisov G.}, \IN{Sov.Phys.JETP}{56}{1982}{258}.
%%CITATION = SPHJA,56,258;%%

\bibitem{Guth:1982ec}
\NAME{Guth A.~H. \atque Pi S.}, \IN{Phys. Rev. Lett.}{49}{1982}{1110}.
%%CITATION = PRLTA,49,1110;%%

\bibitem{Hawking:1982cz}
\NAME{Hawking S.}, \IN{Phys. Lett.B}{115}{1982}{295}.
%%CITATION = PHLTA,B115,295;%%

\bibitem{Bardeen:1983qw}
\NAME{Bardeen J.~M., Steinhardt P.~J. \atque Turner M.~S.}, \IN{Phys.
  Rev.D}{28}{1983}{679}.
%%CITATION = PHRVA,D28,679;%%

\bibitem{Martin:2004um}
\NAME{Martin J.}, \IN{Lect. Notes Phys.}{669}{2005}{199}, [,199(2004)].
%%CITATION = HEP-TH/0406011;%%

\bibitem{Martin:2007bw}
\NAME{Martin J.}, \IN{Lect. Notes Phys.}{738}{2008}{193}.
%%CITATION = ARXIV:0704.3540;%%

\bibitem{Ade:2013zuv}
\NAME{Ade P. \etal}, \IN{Astron.Astrophys.}{571}{2014}{A16}.
%%CITATION = ARXIV:1303.5076;%%

\bibitem{Planck:2013jfk}
\NAME{Ade P. A.~R. \etal}, \IN{Astron. Astrophys.}{571}{2014}{A22}.
%%CITATION = ARXIV:1303.5082;%%

\bibitem{Ade:2015xua}
\NAME{Ade P. A.~R. \etal}, \IN{Astron. Astrophys.}{594}{2016}{A13}.
%%CITATION = ARXIV:1502.01589;%%

\bibitem{Ade:2015lrj}
\NAME{Ade P. A.~R. \etal}, \IN{Astron. Astrophys.}{594}{2016}{A20}.
%%CITATION = ARXIV:1502.02114;%%

\bibitem{Martin:2015dha}
\NAME{Martin J.}, \IN{}{}{2015}{}.
%%CITATION = ARXIV:1502.05733;%%

\bibitem{Schwarz:2001vv}
\NAME{Schwarz D.~J., Terrero-Escalante C.~A. \atque Garcia A.~A.}, \IN{Phys.
  Lett.B}{517}{2001}{243}.
%%CITATION = ASTRO-PH/0106020;%%

\bibitem{Leach:2002ar}
\NAME{Leach S.~M., Liddle A.~R., Martin J. \atque Schwarz D.~J.}, \IN{Phys.
  Rev.D}{66}{2002}{023515}.
%%CITATION = ASTRO-PH/0202094;%%

\bibitem{Liddle:1994dx}
\NAME{Liddle A.~R., Parsons P. \atque Barrow J.~D.}, \IN{Phys.
  Rev.D}{50}{1994}{7222}.
%%CITATION = ASTRO-PH/9408015;%%

\bibitem{Turner:1983he}
\NAME{Turner M.~S.}, \IN{Phys. Rev.D}{28}{1983}{1243}.
%%CITATION = PHRVA,D28,1243;%%

\bibitem{Traschen:1990sw}
\NAME{Traschen J.~H. \atque Brandenberger R.~H.}, \IN{Phys.
  Rev.D}{42}{1990}{2491}.
%%CITATION = PHRVA,D42,2491;%%

\bibitem{Kofman:1997yn}
\NAME{Kofman L., Linde A.~D. \atque Starobinsky A.~A.}, \IN{Phys.
  Rev.D}{56}{1997}{3258}.
%%CITATION = HEP-PH/9704452;%%

\bibitem{Amin:2014eta}
\NAME{Amin M.~A., Hertzberg M.~P., Kaiser D.~I. \atque Karouby J.}, \IN{Int. J.
  Mod. Phys.D}{24}{2014}{1530003}.
%%CITATION = ARXIV:1410.3808;%%

\bibitem{Martin:2006rs}
\NAME{Martin J. \atque Ringeval C.}, \IN{JCAP}{0608}{2006}{009}.
%%CITATION = ASTRO-PH/0605367;%%

\bibitem{Martin:2010kz}
\NAME{Martin J. \atque Ringeval C.}, \IN{Phys. Rev.D}{82}{2010}{023511}.
%%CITATION = ARXIV:1004.5525;%%

\bibitem{Martin:2014nya}
\NAME{Martin J., Ringeval C. \atque Vennin V.}, \IN{Phys. Rev.
  Lett.}{114}{2015}{081303}.
%%CITATION = ARXIV:1410.7958;%%

\bibitem{Martin:2016oyk}
\NAME{Martin J., Ringeval C. \atque Vennin V.}, \IN{Phys.
  Rev.D}{93}{2016}{103532}.
%%CITATION = ARXIV:1603.02606;%%

\bibitem{Mukhanov:1990me}
\NAME{Mukhanov V.~F., Feldman H. \atque Brandenberger R.~H.}, \IN{Phys.
  Rept.}{215}{1992}{203}.
%%CITATION = PRPLC,215,203;%%

\bibitem{Lvovsky:2014sxa}
\NAME{Lvovsky A.~I.}, \IN{}{}{2014}{}.
%%CITATION = ARXIV:1401.4118;%%

\bibitem{Grishchuk:1990bj}
\NAME{Grishchuk L. \atque Sidorov Y.}, \IN{Phys. Rev.D}{42}{1990}{3413}.
%%CITATION = PHRVA,D42,3413;%%

\bibitem{Grishchuk:1992tw}
\NAME{Grishchuk L., Haus H. \atque Bergman K.}, \IN{Phys.
  Rev.D}{46}{1992}{1440}.
%%CITATION = PHRVA,D46,1440;%%

\bibitem{Martin:2012pea}
\NAME{Martin J., Vennin V. \atque Peter P.}, \IN{Phys.
  Rev.D}{86}{2012}{103524}.
%%CITATION = ARXIV:1207.2086;%%

\bibitem{Martin:2015qta}
\NAME{Martin J. \atque Vennin V.}, \IN{Phys. Rev.D}{93}{2016}{023505}.
%%CITATION = ARXIV:1510.04038;%%

\bibitem{Martin:2016tbd}
\NAME{Martin J. \atque Vennin V.}, \IN{Phys. Rev.A}{93}{2016}{062117}.
%%CITATION = ARXIV:1605.02944;%%

\bibitem{Martin:2016nrr}
\NAME{Martin J. \atque Vennin V.}, \IN{Phys. Rev.A}{94}{2016}{052135}.
%%CITATION = ARXIV:1611.01785;%%

\bibitem{Martin:2017zxs}
\NAME{Martin J. \atque Vennin V.}, \IN{Phys. Rev.D}{96}{2017}{063501}.
%%CITATION = ARXIV:1706.05001;%%

\bibitem{Martin:2018zbe}
\NAME{Martin J. \atque Vennin V.}, \IN{JCAP}{1805}{2018}{063}.
%%CITATION = ARXIV:1801.09949;%%

\bibitem{Martin:2018lin}
\NAME{Martin J. \atque Vennin V.}, \IN{}{}{2018}{}.
%%CITATION = ARXIV:1805.05609;%%

\bibitem{Casadio:2004ru}
\NAME{Casadio R., Finelli F., Luzzi M. \atque Venturi G.}, \IN{Phys.
  Rev.D}{71}{2005}{043517}.
%%CITATION = GR-QC/0410092;%%

\bibitem{Casadio:2005xv}
\NAME{Casadio R., Finelli F., Luzzi M. \atque Venturi G.}, \IN{Phys.
  Lett.B}{625}{2005}{1}.
%%CITATION = GR-QC/0506043;%%

\bibitem{Casadio:2005em}
\NAME{Casadio R., Finelli F., Luzzi M. \atque Venturi G.}, \IN{Phys.
  Rev.D}{72}{2005}{103516}.
%%CITATION = GR-QC/0510103;%%

\bibitem{Gong:2001he}
\NAME{Gong J.-O. \atque Stewart E.~D.}, \IN{Phys. Lett.B}{510}{2001}{1}.
%%CITATION = ASTRO-PH/0101225;%%

\bibitem{Choe:2004zg}
\NAME{Choe J., Gong J.-O. \atque Stewart E.~D.}, \IN{JCAP}{0407}{2004}{012}.
%%CITATION = HEP-PH/0405155;%%

\bibitem{Lorenz:2008et}
\NAME{Lorenz L., Martin J. \atque Ringeval C.}, \IN{Phys.
  Rev.D}{78}{2008}{083513}.
%%CITATION = ARXIV:0807.3037;%%

\bibitem{Martin:2013uma}
\NAME{Martin J., Ringeval C. \atque Vennin V.}, \IN{JCAP}{1306}{2013}{021}.
%%CITATION = ARXIV:1303.2120;%%

\bibitem{Ade:2015tva}
\NAME{Ade P. A.~R. \etal}, \IN{Phys. Rev. Lett.}{114}{2015}{101301}.
%%CITATION = ARXIV:1502.00612;%%

\bibitem{Martin:2014lra}
\NAME{Martin J., Ringeval C., Trotta R. \atque Vennin V.}, \IN{Phys.
  Rev.D}{90}{2014}{063501}.
%%CITATION = ARXIV:1405.7272;%%

\bibitem{Gangui:1993tt}
\NAME{Gangui A., Lucchin F., Matarrese S. \atque Mollerach S.}, \IN{Astrophys.
  J.}{430}{1994}{447}.
%%CITATION = ASTRO-PH/9312033;%%

\bibitem{Gangui:1994yr}
\NAME{Gangui A.}, \IN{Phys. Rev.D}{50}{1994}{3684}.
%%CITATION = ASTRO-PH/9406014;%%

\bibitem{Gangui:1999vg}
\NAME{Gangui A. \atque Martin J.}, \IN{Mon. Not. Roy. Astron.
  Soc.}{313}{2000}{323}.
%%CITATION = ASTRO-PH/9908009;%%

\bibitem{Maldacena:2002vr}
\NAME{Maldacena J.~M.}, \IN{JHEP}{05}{2003}{013}.
%%CITATION = ASTRO-PH/0210603;%%

\bibitem{Wands:2007bd}
\NAME{Wands D.}, \IN{Lect. Notes Phys.}{738}{2008}{275}.
%%CITATION = ASTRO-PH/0702187;%%

\bibitem{ArmendarizPicon:1999rj}
\NAME{Armendariz-Picon C., Damour T. \atque Mukhanov V.~F.}, \IN{Phys.
  Lett.B}{458}{1999}{209}.
%%CITATION = HEP-TH/9904075;%%

\bibitem{Garriga:1999vw}
\NAME{Garriga J. \atque Mukhanov V.~F.}, \IN{Phys. Lett.B}{458}{1999}{219}.
%%CITATION = HEP-TH/9904176;%%

\bibitem{Lorenz:2008je}
\NAME{Lorenz L., Martin J. \atque Ringeval C.}, \IN{Phys.
  Rev.D}{78}{2008}{063543}.
%%CITATION = ARXIV:0807.2414;%%

\bibitem{Starobinsky:1992ts}
\NAME{Starobinsky A.~A.}, \IN{JETP Lett.}{55}{1992}{489}, [Pisma Zh. Eksp.
  Teor. Fiz.55,477(1992)].
%%CITATION = JTPLA,55,489;%%

\bibitem{Hazra:2010ve}
\NAME{Hazra D.~K., Aich M., Jain R.~K., Sriramkumar L. \atque Souradeep T.},
  \IN{JCAP}{1010}{2010}{008}.
%%CITATION = ARXIV:1005.2175;%%

\bibitem{Martin:2011sn}
\NAME{Martin J. \atque Sriramkumar L.}, \IN{JCAP}{1201}{2012}{008}.
%%CITATION = ARXIV:1109.5838;%%

\bibitem{Hazra:2012yn}
\NAME{Hazra D.~K., Sriramkumar L. \atque Martin J.},
  \IN{JCAP}{1305}{2013}{026}.
%%CITATION = ARXIV:1201.0926;%%

\bibitem{Martin:2014kja}
\NAME{Martin J., Sriramkumar L. \atque Hazra D.~K.},
  \IN{JCAP}{1409}{2014}{039}.
%%CITATION = ARXIV:1404.6093;%%

\bibitem{Avila:2013ela}
\NAME{Ávila S., Martin J. \atque Steer D.}, \IN{JCAP}{1408}{2014}{032}.
%%CITATION = ARXIV:1304.3262;%%

\bibitem{Martin:2013tda}
\NAME{Martin J., Ringeval C. \atque Vennin V.}, \IN{Phys. Dark
  Univ.}{5-6}{2014}{75–235}.
%%CITATION = ARXIV:1303.3787;%%

\bibitem{Martin:2013nzq}
\NAME{Martin J., Ringeval C., Trotta R. \atque Vennin V.},
  \IN{JCAP}{1403}{2014}{039}.
%%CITATION = ARXIV:1312.3529;%%

\bibitem{Trotta:2008qt}
\NAME{Trotta R.}, \IN{Contemp. Phys.}{49}{2008}{71}.
%%CITATION = ARXIV:0803.4089;%%

\end{thebibliography}

\end{document}